\documentclass{jltp}
\usepackage{graphicx}
\usepackage{psfig,graphics}
\usepackage{color}
\usepackage{amssymb}
\usepackage{bookmath}
\def\vec#1{{\mathbf{#1}}}
\title{Spectrum of Third Sound Cavity Modes on Superfluid \He\ Films}
\author{A. Vorontsov and J. A. Sauls}
\address{Department of Physics and Astronomy, \\ 
         Northwestern University, Evanston, IL 60208, USA}
\runninghead{A. Vorontsov and J. A. Sauls}{Third Sound Cavity Modes}
\begin{document}
%----------------------------------------------------------------------------------
\maketitle
\begin{abstract}
We report theoretical calculations of the spectrum
of third sound modes for a cylindrically symmetric
film of superfluid \He, and compare these results
with experimental data for the mode frequencies and
amplitude spectrum of surface waves of superfluid
\He\ films.
\end{abstract}
\section{INTRODUCTION}
\begin{figure}
\centerline{\includegraphics[height=3cm]{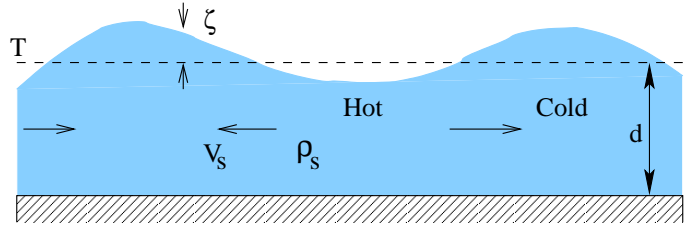}}
\caption{Third sound oscillations of the superfluid produce
         surface waves.}
\label{fig:geometry}
\end{figure}

Third sound is an oscillation of the superfluid component in the plane of the
film, with the normal component clamped by viscosity to the substrate
that supports the film. This
sound mode was first discussed theoretically by Atkins for superfluid
\Hefour\ films in the context of the two-fluid model.\cite{atk59}
The oscillations of the superfluid component
cause the level of the film to locally rise and fall
(Fig.~\ref{fig:geometry}).
The restoring force that stabilizes the
oscillations is the van der Waals attraction between the helium and the
substrate. The possibility of observing third sound in superfluid \He\
films was discussed by Eggenkamp et al.,\cite{egg98} and the first
experimental report of surface modes identified as third sound was by
Schechter et al.\cite{sch98} Here we report theoretical calculations of
the spectrum of cavity modes of third sound in superfluid \He\ films
with cylindrical symmetry and compare the results with the observed
spectra.

%~~~~~~~~~~~~~~~~~~~~~~~~~~~~~~~~~~~~~~~~~~~~~~~~~~~~~~~~~~~~~~~~~
%~~~~~~~~~~~~~~~~~~~~~~~~~~~~~~~~~~~~~~~~~~~~~~~~~~~~~~~~~~~~~~~~~

\section{CAVITY MODES}

We consider \He\ films with thicknesses of order $d\simeq
300-400\,\mbox{nm}$, roughly $3-6$ coherence lengths at zero pressure. These
are thin films in the sense that they are expected to be in the either
the axial or planar phase.\cite{vor03} In either case the orbital anisotropy
axis, $\vell$, is anchored normal to the substrate. The analysis
reported here assumes that the $\vell$ vector remains fixed in the presence
of small oscillations of the surface. We expect this to be a good
approximation for long-wavelength, low-frequency third sound, since the
coupling of the low-frequency normal-flapping mode in \He-A to transverse
currents vanishes in the limit of exact particle-hole symmetry.\cite{yip92}
We also neglect edge currents associated with the broken chiral symmetry of 
\He-A, which may play an important role in nearly two-dimensional films with
specular surfaces, but are not relevant for third sound in the geometry 
considered here.\cite{vol92}
Thus, in zero magnetic field the hydrodynamics of the superfluid film is
defined by the phase mode and the surface excitation of the film, and reduces
to scalar hydrodynamics. At temperatures $T\sim\mbox{mK}$ the viscous
penetration depth is of order $\delta=\sqrt{2\eta/\rho\omega}\approx
1\,\mbox{cm}\gg d$, for low frequency third sound ($\nu \sim 1\,$Hz)
($\eta$ and $\rho$ are the viscosity and density of normal \He),
and so we can safely consider the excitations (normal component) to be 
viscously clamped to the substrate.
Furthermore, for third sound, the motion of the condensate is
incompressible and confined to the plane of the film. Oscillations of the
condensate then lead to height oscillations of the film.
The hydrodynamic equations governing the excitation of the film
include the continuity equation for mass conservation,
%---------------------------------------------------------------------------------
\be
\rho{\partial\zeta\over \partial t}+\rho_s d\,\grad\cdot\vec{v}_s=0
\,,
\label{eq:continuity}
\ee
%---------------------------------------------------------------------------------
where $\rho$ is the mass density, $\zeta$ is the deviation of the film height
from its equilibrium value, $d$, and $\rho_s$ is the superfluid density
averaged over the thickness of the film. The pair momentum, or superfluid
velocity, related to the phase $\vartheta$ of the condensate's wave function,
$\vec{p}_s=\frac{\hbar}{2}\grad\vartheta=m_3\vec{v}_s$, describes
the in-plane motion of the condensate and obeys the Josephson equation of motion,
%---------------------------------------------------------------------------------
\be
{\partial \vec{v}_s\over\partial t}=-\grad\mu=-f\grad\zeta+s_f\grad T
\,,
\label{eq:josephson}
\ee
%---------------------------------------------------------------------------------
where $\mu$ is the chemical potential,
$f$ is the van der Waals force per unit mass, $s_f$ is the entropy per unit mass 
and $T$ is the absolute temperature.
The oscillations of the condensate are coupled to entropy oscillations, and thus
the height oscillations of third sound are coupled to the heat current in the film.
This coupling, described by the heat transport equation,
%---------------------------------------------------------------------------------
\be
\rho c_f d\pder{T}{t}+\rho s_f T\pder{\zeta}{t}-d\kappa_f\grad^2 T = 0
\,,
\label{eq:heat}
\ee
%---------------------------------------------------------------------------------
where $c_f$ is the specific heat and $\kappa_f$ is the thermal conductivity per
unit mass, is a key mechanism for damping of third sound cavity modes. 

If we neglect heat transport in the film, then the height deviation,
$\zeta$, and the superfluid velocity, $\vec{v}_s$, obey two-dimensional wave
equations with the phase velocity given by $c=\sqrt{\rho_s(T) d f/\rho}$.
For cylindrically symmetric films the eigenfunctions are products of
Bessel functions (regular at $r=0$) and azimuthal harmonics,
$J_m(kr)\,e^{im\phi}$, where $k=\omega/c$ is the wavevector for the
mode and $m=0,\pm 1, \pm 2, \ldots$. The cavity mode eigenfrequencies
are determined by the boundary conditions at the edge ($r=R$) of the
film. For fixed boundary conditions, $\zeta(R)=0$, the resonant
frequencies and wavevectors are $\omega_{mn} = c(T)\, a_{mn}/R$,
where $a_{mn}$ is the
$n^{\mbox{\tiny th}}$ zero of $J_m(x)$. For free boundary conditions,
$\partial_r\zeta(R)=0$, $\omega_{mn}=c(T)\,a'_{mn}/R$, where
$(a'_{mn})$ are the zeroes of the derivative, $J_m'(x)$. The
general solution to the wave equation can then be expressed as an
eigenmode expansion,
%---------------------------------------------------------------------
\be
\zeta(r,\phi,t)=\sum_{m,n}A_{mn}\;J_m(\omega_{mn}r/c)\,
                          e^{im\phi}\;e^{-i\omega_{mn}\,t}
\,,
\ee
%---------------------------------------------------------------------
where the amplitudes, $A_{mn}$, are determined by driving forces that
couple to the displacement.

\section{ANALYSIS}

\begin{figure}
\centerline{\includegraphics[width=8.0cm,height=10cm,angle=-90]{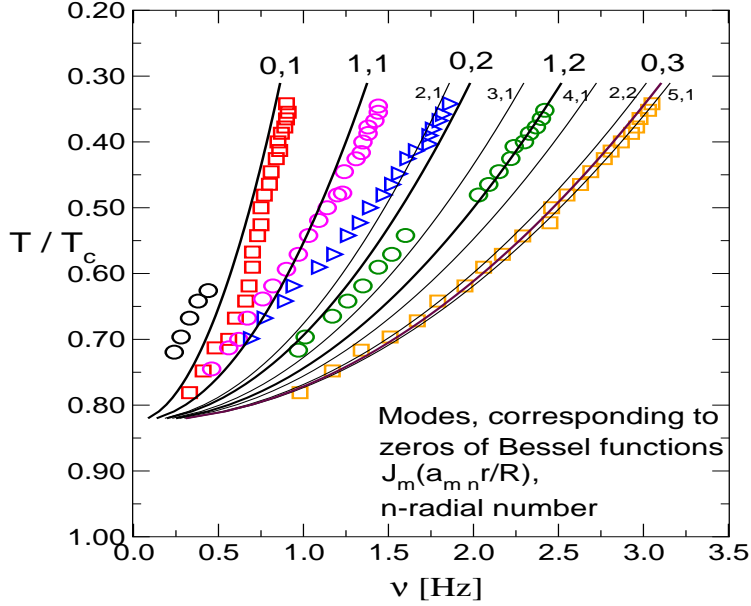}}
\caption{\label{f:ident} Mode frequencies for a $250$ nm film (symbols) from 
Ref. \onlinecite{sch99}. The
solid curves are the calculated mode frequencies with mode indices as labelled
for fixed boundary conditions.}
\end{figure}

Schechter et al.\cite{sch98} observed standing waves of third sound on a
circular substrate by exciting the surface with an a.c. electric field
normal to the film. As noted in Ref. \onlinecite{sch99} the modes for
thicker films ($250-400$ nm) are reasonably well indexed by assuming fixed
boundary conditions, whereas thinner films ($d<250$ nm) are better indexed
assuming a free boundary condition. The mode frequencies for a $250$ nm
thick film are shown in Fig.~\ref{f:ident}. The temperature dependence
reflects that of the superfluid density, which is approximately described by 
the Ginzburg-Landau form, $\rho_s(T)/\rho\propto 1-T/T_c^{\mbox{\tiny film}}$.
The fit of the theoretical mode frequencies to the experimental data is done 
assuming that only the lowest frequency modes with
$m=0,1$ are detected. In general there are higher frequency modes that may
contribute, as well as modes with $m\ge 2$; these modes are also shown in
Fig.~\ref{f:ident} for comparison.
%~~~~~~~~~~~~~~~~~~~~~~~~~~~~~~~~~~~~~~~~~~~~~~~~~~~~~~~~~~~~~~~~~~~~~~~~~~~~~
%~~~~~~~~~~~~~~~~~~~~~~~~~~~~~~~~~~~~~~~~~~~~~~~~~~~~~~~~~~~~~~~~~~~~~~~~~~~~~
It is not {\sl apriori} obvious that these modes can be ignored.
There are significant differences between theory and 
experiment, particularly for the lowest order modes.
In order to more precisely identify the modes and to test the hydrodynamic
theory for third sound oscillations of \He\ we have calculated the amplitude
response of the film. The calculation includes the sensitivity of the detector 
for the various modes, and provides a determination of the amount of energy 
that is pumped into each mode.

The excitation field, or driving force, enters via the electrostatic energy density
of the thin film which is a weak dielectric. The electro-chemical potential is then
%---------------------------------------------------------------------------------
\be
\mu=\mu_0(T,\rho)-{\vec{E}^2 \over 8 \pi}{\epsilon-1 \over \rho}
\,,
\ee
%---------------------------------------------------------------------------------
where $\rho$ is the mass density, $\epsilon$ is the dielectric constant of the
liquid \He\ and $\vec{E}$ is the electric field. 
The response of the film to an a.c. voltage, $V(t)=V_0+V_1 e^{i\omega t}$, 
is governed by the wave equation with the driving force on the right-hand side,
%---------------------------------------------------------------------------------
\be
{\partial^2 \zeta \over \partial t^2}-c^2 \nabla^2\zeta =
-c^2(T) \, A \, \sum_{m,n} f_{mn} J_m(a_{mn}{r\over R}) \cos(m\phi)
\left( 2 V_1\over V_0 \right) e^{i\omega\, t}
\ee
%---------------------------------------------------------------------------------
where $A\sim 2 \mbox{\AA}$ is of the order of the surface wave amplitude and
$f_{mn}$ is the strength of the excitation force for mode $(m,n)$. 
The capacitive excitation ring is displaced above the film and is slightly tilted 
about an axis (y-axis) the horizontal plane of the film.\cite{sch98} If
we measure the azimuthal angle with respect to the orthogonal direction in
the plane (x-axis) then the 
resulting forces on the film are symmetric with respect to $\phi\to -\phi$.
The surface displacement is then
%---------------------------------------------------------------------------------
\be
\zeta(r,\phi, t) = A\, \left( 2 V_1\over V_0 \right) c^2
\sum_{m,n} {f_{mn} \over \omega_{mn}^2-\omega^2}
J_m(a_{mn}r/R)\cos (m\phi)\,e^{i\omega\,t}
\,.
\ee
%---------------------------------------------------------------------------------

\begin{figure}
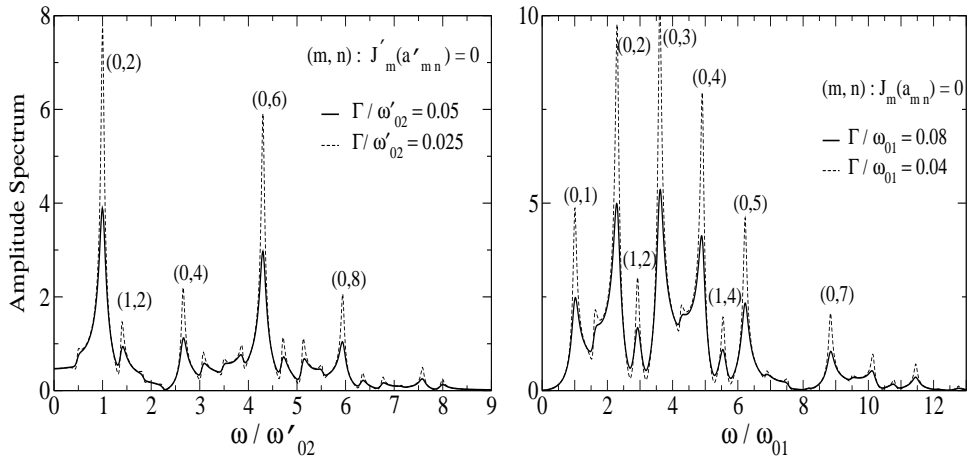

\hspace*{0.5cm}
\begin{minipage}{5cm}
\centerline{\includegraphics[width=6.5cm,height=6cm]{./pictres1aps.eps}}
\end{minipage}
\hspace{1.0cm}
\begin{minipage}{5.5cm}
\centerline{\includegraphics[width=6.0cm,height=6cm]{./pictres2aps.eps}}
\end{minipage}
\caption{\label{f:damping} Spectral weight for various modes of third sound
folded with the detector sensitivity. The spectra are calculated assuming a
constant, frequency independent damping, $\Gamma$, with free boundary conditions \
(left) and fixed boundary conditions (right).}
\end{figure}

The capacitive detection method used in Refs. \onlinecite{sch98,sch99} 
allows one to measure the average
displacement of the film surface. The amplitude spectrum,
$A(\omega)$, is proportional to the average displacement and is given by
%---------------------------------------------------------------------------------
\be\label{eq:amplitude_spectrum}
A(\omega)=A\, \left( 2 V_1\over V_0 \right) \, c^2(T) \,
\left| \sum_{m,n} {F_{mn} \over \omega_{mn}^2-\omega^2} \right|
\,,
\ee
%---------------------------------------------------------------------------------
where $F_{mn}$ is the product of the excitation force ($f_{mn}$) and
the detector sensitivity ($D_{mn}$) for mode $(m,n)$: $F_{mn}
= f_{mn}\times\,D_{mn}$.
The detector sensitivity is determined by the electric field distribution
for the geometry of the capacitor, which we calculate for the experimental 
setup of Ref. \onlinecite{sch98}.
The modes shown in Fig.~\ref{f:damping} for the response
are indeed the $m=0,1$ modes, which justifies the earlier assumption.
%---------------------------------------------------------------------------------
%---------------------------------------------------------------------------------
However, the spectra shown in Fig.~\ref{f:damping} differ significantly from the experimental
spectra. There are many more modes visible in the
calculated spectrum. This is a consequence of the assumption of frequency independent
damping for the modes; constant damping does not suppress the
higher frequency modes.

\begin{figure}
\begin{minipage}{5cm}
\centerline{\includegraphics[width=6cm,height=6cm]{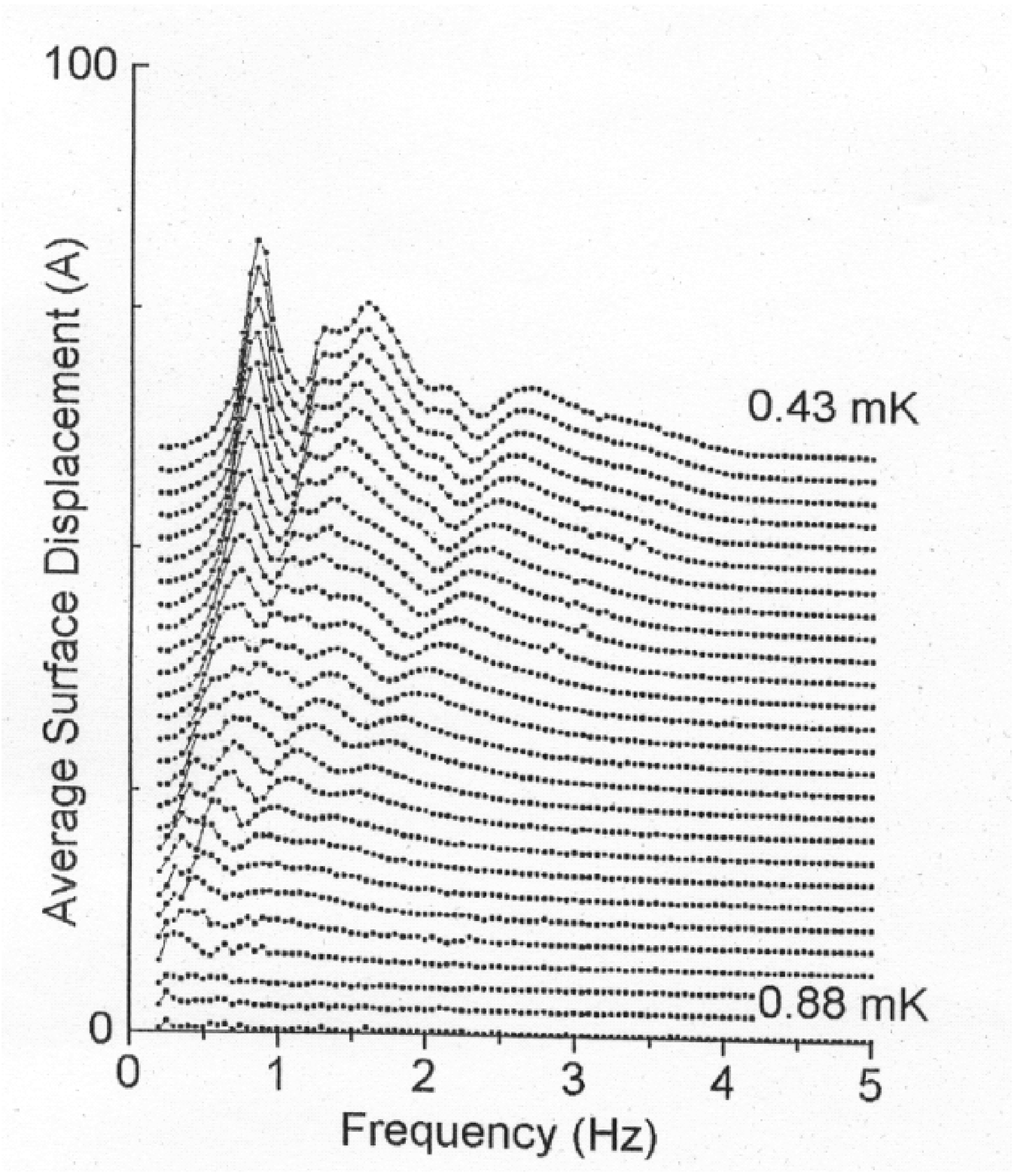}}
\end{minipage}
\hspace{0.7cm}
\begin{minipage}{6cm}
\centerline{\includegraphics[width=7cm,height=6cm]{./pictres4aps.eps}}
\end{minipage}
\caption{\label{f:fixed} Experimental spectra from Ref. \onlinecite{sch99} for a
``thick'' film  with $d=380$ nm (left) and the corresponding theoretical spectra
calculated for fixed boundary conditions. Spectra for different temperatures are
shifted vertically as indicated in the experimental figure.}
\end{figure}

\begin{figure}
\begin{minipage}{5cm}
\centerline{\includegraphics[width=5.5cm,height=6cm]{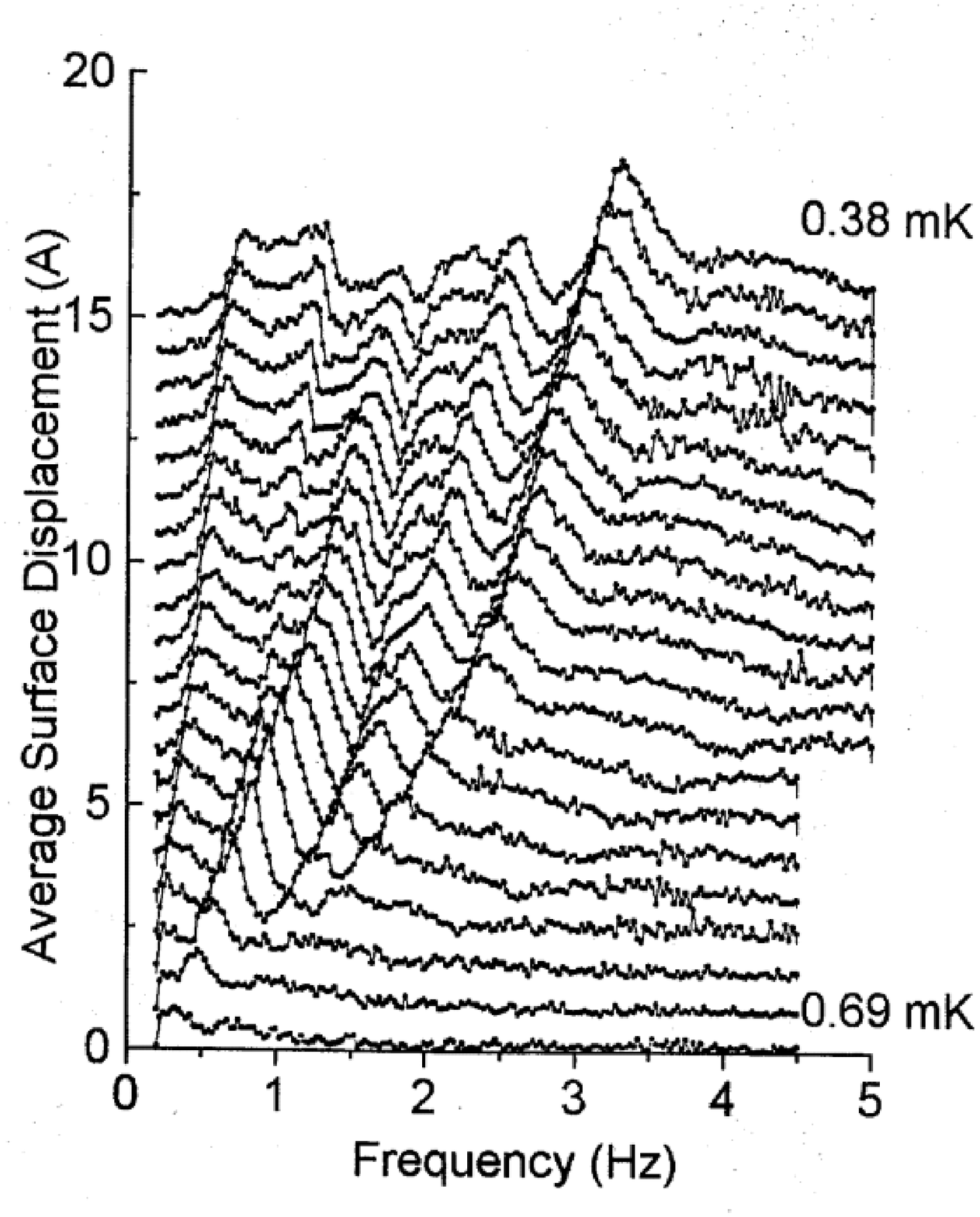}}
\end{minipage}
\hspace{0.5cm}
\begin{minipage}{6cm}
\centerline{\includegraphics[width=7cm,height=6cm]{./pictres3aps.eps}}
\end{minipage}
\caption{\label{f:free} Experimental spectra from Ref. \onlinecite{sch99}
for a ``thin'' film (230 nm) and the corresponding
theoretical spectra calculated with free boundary conditions.}
\end{figure}

In general the damping of third sound is frequency dependent. In the following we calculate
the frequency dependent damping rate for third sound that results from heat transport
induced by the oscillations of the entropy associated with oscillations of the condensate.
The resulting amplitude spectrum is obtained by replacing
$\omega\to\omega +i\Gamma(T,\omega)$ in Eq. \ref{eq:amplitude_spectrum}
in the response formula. The damping rate contains two terms,
$\Gamma=\Gamma_0+\Gamma_1(T,\omega)$,
where $\Gamma_0$ is a constant damping rate which we attribute to roughness of the substrate
and $\Gamma_1(\omega, T)$ is the damping obtained from the heat transport equation
(Eq. \ref{eq:heat}),
%---------------------------------------------------------------------------------
\be
\Gamma_1(\omega, T) =
{1\over2} \left(\frac{s_f}{c_f}\right)^2
\frac{T \kappa_f(T)}{f d \rho  c^2(T)} \, \omega^2
\equiv \Gamma_1 \cdot {t^2\over 1-t} \, \omega^2
\,,
\ee
%---------------------------------------------------------------------------------
where $t = T/T_c^{\mbox{\tiny film}}$. Note that the damping
rate diverges for $T\to T_c^{\mbox{\tiny film}}$ because the
superfluid stiffness vanishes whereas the thermal conductivity
is finite. We have also assumed that surface pair breaking
yields superfluidity with a gapless spectrum, in which case
$s_f=c_f\approx \gamma_s T$ and $\kappa_f\propto T$.\cite{vor03} The
resulting damping rate leads to strong suppression of the higher
frequency modes and much closer agreement with the experimental
spectra. The calculated results for the amplitude spectra with
both fixed and free boundary conditions are shown in comparison
with the experimental spectra in Figs. \ref{f:fixed} and \ref{f:free}.
%---------------------------------------------------------------------------------
%---------------------------------------------------------------------------------
For the thick film with $d=380\,\mbox{nm}$ the theoretically calculated
spectrum are in reasonable agreement with the experimental spectrum for
temperatures $T \lesssim 0.64\,\mbox{mK}$. At higher temperatures mixing or
crossing of the modes is seen in the experimental spectrum. Such features are
not observed in the theoretical spectra obtained for either type of boundary
condition. The discrepancies are more severe for thinner films. In this case
neither boundary condition yields agreement between the calculated and
observed spectra. Experimentally, there is a much richer mode structure than
the theoretically predicted spectrum.
%---------------------------------------------------------------------------------
%---------------------------------------------------------------------------------

\section{CONCLUSIONS}

In summary,
the thicker films (250-400 nm) at low temperatures are reasonably well described
by the equations of two-fluid hydrodynamics with damping due to surface
roughness and thermal transport. The latter mechanism suppresses the higher frequency
modes and diverges for $T\to T_{c}^{\mbox{\tiny film}}$. In this regime the
theoretically predicted modes and their spectral weights correspond to those
observed in the experiment. The apparent mode splitting
or mixing that onsets at higher temperatures, $T\gtrsim 0.65 T_{c}^{\mbox{\tiny film}}$,
may indicate a cross-over to a another mechanism for dissipation or, what seems more
likely given the abruptness of the onset, a phase transition of the superfluid phase
of the film. For thinner films (170-250 nm) we are not able to describe the observed
spectra using two-fluid hydrodynamics with either set of boundary conditions and the
same damping mechanisms. Although the frequencies of the
modes can be associated with modes calculated with free boundary conditions,
the spectral weights for the modes calculated from the theory
do not agree with experiment. This fact, and mixing or splitting observed in
thicker films may indicate that there is a phase transition at temperatures
well below $T_c$ for films with $d\sim 3-6\,\xi_0$.

\section*{ACKNOWLEDGEMENTS}
We would like to thank A.~Schechter for making his thesis available to us.
This research was supported by NSF grant DMR-9972087.

\end{document}